\documentclass[12pt]{article}
\usepackage[all]{xy}

\usepackage{etex}
\usepackage{amsmath,amssymb,amscd,bm,mathtools,array}

\usepackage[margin=1.75cm]{caption} 
\usepackage[labelfont=it,textfont={small,it}]{caption} 

\usepackage{color}
\usepackage[colorlinks=true, pdfstartview=FitV, linkcolor=blue, citecolor=blue, urlcolor=blue]{hyperref}

\usepackage[cyr]{aeguill}     
\usepackage[pdftex]{graphicx} 
\DeclareGraphicsExtensions{.jpg,.mps,.pdf,.png} 

\let\ssection=\section
\renewcommand{\section}{\setcounter{equation}{0}\ssection}

\newcommand{\cE}{{\mathcal{E}}}

\newcommand{\rg}{\mathrm{g}}

\newcommand{\cL}{{\mathcal{L}}}

\newcommand{\bcL}{\bm{\cL}}

\newcommand{\rO}{\mathrm{O}}

\newcommand{\so}{\mathfrak{so}}
\newcommand{\cO}{{\mathcal{O}}}

\newcommand{\bomega}{\boldsymbol{\omega}}

\newcommand{\barP}{\overline{P}}

\newcommand{\bp}{{\mathbf{p}}}

\newcommand{\np}{\Vert\bp\Vert}

\newcommand{\Pf}{\mathrm{Pf}}
\newcommand{\dP}{\dot{P}}

\newcommand{\bbR}{\mathbb{R}}

\newcommand{\bs}{{\mathbf{s}}}
\newcommand{\bst}{{\mathbf{s^\bot}}}
\newcommand{\ns}{\Vert\bs\Vert}
\newcommand{\nst}{\Vert\bst\Vert}

\newcommand{\sign}{\mathrm{sign}}

\newcommand{\dS}{\dot{S}}

\newcommand{\Tr}{\mathrm{Tr}}

\newcommand{\bu}{{\mathbf{u}}}

\newcommand{\bv}{{\mathbf{v}}}

\newcommand{\bw}{{\mathbf{w}}}

\newcommand{\dX}{\dot{X}}
\newcommand{\bx}{{\mathbf{x}}}

\newcommand{\xps}{\widetilde{\mathrm{xps}}}
\newcommand{\xl}{\widetilde{\mathrm{x\cL}}}

\newcommand{\half}{\frac{1}{2}}

\newcommand{\lb}{\left[}
\newcommand{\rb}{\right]}
\newcommand{\lp}{\left(}
\newcommand{\rp}{\right)}
\newcommand{\bb}{\begin{eqnarray}}
\newcommand{\ee}{\end{eqnarray}}
\newcommand{\eee}{\nonumber\end{eqnarray}}
\newcommand{\qq}{\quad}
\newcommand{\dpp}{\vcentcolon}

\newcommand{\arcsec}{^{\prime\prime}}

\def\b{\beta}

\def\d{\delta}

\begin{document}

\thispagestyle{empty}

\begin{center}
${}$
\vspace{2cm}

{\Large\textbf{Gravitational birefringence of light\\
in Schwarzschild spacetime\footnote{supported by the OCEVU Labex (ANR-11-LABX-0060) funded by the
"Investissements d'Avenir" 
\\\indent\qq
French government program
}}} \\

\vspace{2cm}

{\large
Christian Duval, Lo\"ic Marsot\footnote{mailto: marsot.loic@gmail.com}
and 
Thomas Sch\"ucker\footnote{mailto: thomas.schucker@gmail.com}
\\[8pt]
Aix Marseille Univ, Universit\'e de Toulon, CNRS, CPT,
Marseille, France }

\vspace{2cm}

{\large\textbf{Abstract}}

\end{center}

We compute the gravitational birefringence of light as it undergoes gravitational lensing. To this end we re-derive the Souriau-Saturnini equations in the Schwarzschild metric and solve them numerically and perturbatively. Our main result is an offset between the trajectories of the photons of opposite polarisations, which grows  with time. We also find an intriguing instability of the spin component transverse to the momentum. 

\vspace{3cm}

\hfill {\em To the memory of Christian Duval}

\vspace{2cm}

\vskip 1truecm

\eject

\section{Introduction}

Birefringence of light is a well kown phenomenon in anisotropic matter like quartz or calcite. On theoretical grounds Federov \cite{Fed55} in 1955 and Imbert \cite{Imb72} in 1972 predicted birefringence in absence of matter but in an electric field with a gradient. This prediction was confirmed experimentally in 2008 by Bliokh \cite{BNKH08} et al. and Hosten \& Kwiat \cite{HK08}. They measured an offset between the trajectories of the photons of opposite polarisations, which is of the order of the wavelength of the photons.

Birefringence of light without matter arises in Loop Quantum Gravity \cite{gp} and more generally in extensions of Maxwell's theory or of the $SU(3)\times SU(2) \times U(1)$ model, which violate Lorentz invariance, like the addition of a Chern-Simons term, \cite{ck,jk}. Birefringence in these theories is strongly constrained by spectropolarimetry of radio galaxies, quasars, Gamma Ray Bursts and the Cosmic Microwave Background \cite{cfj,km01,km06,km07,km13}.

The starting point of the present work is less ambitious: the Mathisson-Papapetrou-Dixon equations \cite{Mat37,Papa51,Dix70} and their geometric derivation in 1974 by Souriau \cite{Sou74}. He and Saturnini~\cite{Sat76}, using this geometric framework,  generalized the null geodesic equation in order to include the spin of photons, thus proposing birefringence of light in a gravitational field with a gradient. In 1976 Saturnini computed this generalized equation in the Schwarzschild metric and obtained first numerical solutions with birefringence.

The usual geodesic equation can be derived exactly from general principles, when neglecting dipole and higher moments of the test particle. See the derivation using conserved quantities by Dixon \cite{Dix70} or, more geometrically by Souriau \cite{Sou74}. This computation can be generalized to include the dipole moment of the test particle. This is how the Mathisson-Papapetrou-Dixon~(MPD) equations are derived. They can also be obtained using Souriau's geometrical approach. With $X=(X^\mu(\tau))$ denoting the trajectory of the test particle,~$P$ its 4-momentum, and $S$ a skewsymmetric tensor describing its spin state, we have the MPD equations,
\begin{align}
\label{mpd}
\dot{P}^\mu & = - \half {R^\mu}_{\rho\alpha\beta} S^{\alpha \beta} \dot{X}^\rho, \\
\dot{S}^{\mu\nu} & = P^\mu \dot{X}^\nu - P^\nu \dot{X}^\mu,
\end{align}
where the dot over the trajectory $X$ denotes the ordinary derivative with respect to its affine parameter, $\dot{X} = dX/d\tau$, while the dot over $P$ and $S$ denotes the covariant derivative with respect to that same parameter.

These equations are very general and can be applied to any kind of test particle, be it a planet with its intrinsique angular momentum or an elementary particle with its spin. However, these equations are not deterministic: there are more unknowns than equations. We need to impose constraints, or ``equations of state'', that depend on the test particle we want to describe. In flat spacetime, at least for photons, ${S^\mu}_\nu P^\nu = {S^\mu}_\nu \dot{X}^\nu = 0$ holds. Souriau shows \cite{Sou70} that the MPD equations together with the previous constraint in flat spacetime for massless particles of spin~1 lead to the Maxwell equations after geometric quantization. Now, in curved spacetime the MPD equations entail that the 4-momentum~$P$ may not be parallel to the 4-velocity $\dot{X}$, thus there are different possible equations of state. For extended particles, it is generally accepted to impose the Tulczyjew constraint, ${S^\mu}_\nu P^\nu = 0$ \cite{Tul59}, to describe uniquely the worldline of the center of mass of a particle \cite{Dix70}. However, for elementary particles, the center of mass gives no criterion. An alternative possibility for the equation of state would then be the Frenkel-Pirani constraint  ${S^\mu}_\nu \dot{X}^\nu = 0$ \cite{Fre26,Pir56}. For examples with the Frenkel-Pirani constraint, see \cite{Wey47,Mas75,Duv78,Obu11}.

In this paper, we consider the Tulczyjew constraint since this framework can be successfully applied for elementary particles in different situations: in the electromagnetic field, this constraint yields the well known and well used Bargmann-Michel-Telegdi equations \cite{BMT59}, with an anomalous velocity, and with correct anomalous magnetic moment \cite{Duv16}; it has also been used \cite{Duv06,Duv07,Duv08,Duv13} to correctly describe the above mentioned birefringence in the Fedorov-Imbert effect.

There are also examples in gravitational fields, with massive particles, see \cite{Moh01,Obu11}. Let us insist that solving the massless equations in this framework is more involved than for massive equations.

Reference \cite{rw} presents the Souriau-Saturnini equations in a generic Robertson-Walker metric and some numerical and some perturbative solutions, showing a striking effect on the photon's trajectory. Indeed, it travels on a helix, centered around the null geodesic  and with a radius of the order of the wavelength.

Here we take up Saturnini's work in the (outer) Schwarzschild metric. Thanks to present day computing power, we obtain numerical solutions precise enough to lead us to perturbative solutions. These solutions, contrary to the ones in a Robertson-Walker metric, feature an intriguing instability of the spin component transverse to the momentum and an offset between the trajectories of photons with opposite polarisations, which grows linearly with time.

In 2006 Gosselin,  B\'erard \& Mohrbach \cite{gbm} have published an analysis similar to ours, but starting from the Bargmann-Wigner equations. We compare results at the end of section~\ref{s:5}.

Let us note an experimental upper bound on birefringence obtained in 1974 by Very Long Baseline Interferometry of  radio sources lensed in the Sun's gravitational field \cite{vlbi}.

The paper is organized as follows. We will start with the Souriau-Saturnini equations in section~\ref{s:2}, which are the application of the Mathisson-Papapetrou-Dixon equations with the Tulczyjew constraint to a massless particle of spin 1, in the case of a Schwarzschild spacetime. Section~\ref{s:3} has two aims: to show the similarity between the usual geodesic equations and the equations obtained by taking the photon's spin into consideration; and to present a different approach to the well known gravitational lensing in Schwarzschild spacetime based on the conserved quantities derived here and on first order differential equations. The following section~\ref{s:4} will be devoted to the numerical study of the complicated equations of motion with spin obtained from section~\ref{s:2}. In light of the numerical integrations and the similarity of the spinning equations and the geodesic equations, the last section~\ref{s:5} presents a perturbative solution to the Souriau-Saturnini equations in Schwarzschild spacetime, which will allow us to interpret the differences introduced by the spin of the photon in gravitational lensing.

\section{Spinning massless particles}
\label{s:2}

\subsection{The Souriau-Saturnini equations}

We use the shorthand notations,
\begin{equation}
\label{notations}
{R(S)^\mu}_\nu \dpp= {R^\mu}_{\nu\alpha\beta} S^{\alpha\beta} \qquad \mathrm{and }\qquad R(S)(S) \dpp= R_{\mu\nu\alpha\beta}S^{\mu\nu}S^{\alpha\beta}.
\end{equation}
We also suppress indices by using linear maps, \textit{i.e.} $S = ({S^\mu}_\nu)$, and we write $P = (P^\mu)$.  Assuming the consistency condition $R(S)(S)\neq0$, the equations of motion of photons in space-time~\cite{Sat76} read:
\begin{eqnarray}
\label{dotXter}
\dX&=&P+\frac{2}{R(S)(S)}S R(S) P\,,\\[4pt]
\label{dotPter}
\dP&=&-s\,\frac{\Pf(R(S))}{R(S)(S)}\,P\,,\\[4pt]
\label{dotSter}
\dS&=&P\overline{\dX}-\dX\barP.
\end{eqnarray}
Here $\barP$ denotes the covector associated to the vector $P$ via the metric: $\barP_\mu \dpp=g_{\mu \rho }P^\rho $. We denote the Pfaffian of a skewsymmetric linear map $F$ as $\Pf(F)$. It is the square root of the determinant of the linear map, noting that the determinant of a skewsymmetric matrix can always be written as a perfect square. An alternative definition of the Pfaffian is $\Pf(F)=-\frac{1}{8}\sqrt{-\det(\rg_{\alpha\beta})}\,\varepsilon_{\mu\nu\rho\sigma}F^{\mu\nu}F^{\rho\sigma}$ with $\varepsilon_{\mu\nu\rho\sigma}$ the Levi-Civita symbol such that $\varepsilon_{1234}=1$.

For a derivation of these equations (\ref{dotXter} - \ref{dotSter}) in English, see \cite{rw}.

\subsection{Metric}

The Schwarzschild metric can be expressed in an isotropic coordinate patch $(X^\mu)=(\bx,t)$ by
\begin{equation}
\rg = -B^2\Vert d\bx\Vert^2+A^2\,dt^2
\label{gbis}
\end{equation}
with
\bb
A\dpp=\,\frac{r-a}{r+a}\,, \qq B\dpp=\lp\frac{r+a}{r}\rp^2,\qq r\dpp=\sqrt{\bx\cdot\bx}\,,\qq 0<a<r\,.
\ee
where $\bx=(x^1,x^2,x^3)$ and $\Vert\,\cdot\,\Vert$ is the Euclidean norm.
If $(\rho,\theta,\varphi,t)$ are the Schwarzschild coordinates, the isotropic polar ones $(r,\theta,\varphi,t)$ are related by 
\bb
\rho=r\left(1+\frac{GM}{2r}\right)^2
\qquad
\text{or}
\qquad
r=\half\left(\rho-GM+\sqrt{\rho(\rho-2GM)}\right).
\ee
Recall that $a=\half GM$ is the Schwarzschild radius.

The vector product, which abounds in computations involving spin, take a simple form in isotropic coordinates. Therefore we adopt these coordinates. However they have the drawback, that  including  the cosmological constant, which is straight forward in the Schwarzschild coordinates, becomes difficult. This inclusion will be dealt with in a future publication.

We have the following Christoffel symbols,
\begin{equation}
{\Gamma^j}_{ii}=-{\Gamma^i}_{ji}=-{\Gamma^j}_{jj}=\frac{2a\, x^j}{r^2(r+a)}\,,
\quad
{\Gamma^j}_{44}=\frac{2ar^3(r-a)\, x^j}{(r+a)^7}\,,
\qq
{\Gamma^4}_{4j}=\frac{2a\, x^j}{r\,(r+a)(r-a)}\,,
\label{Gamma}
\end{equation}
for all $i\not=j=1,2,3$, no summation over repeated indices. 

For the Riemann tensor ${R^\mu}_{\nu\alpha\beta}=\partial_{\alpha}{\Gamma^\mu}_{\beta\nu}-\partial_{\beta}{\Gamma^\mu}_{\alpha\nu}+\cdots$ with $i,j$ and $k$ all different, we have
\begin{align}
{R^i}_{jij}&=\,\frac{2a\,[2(x^k)^2-(x^i)^2-(x^j)^2]}{r^3(r+a)^2}\,,&
 {R^j}_{iki}&=-\,\frac{6a\,x^jx^k}{r^3(r+a)^2}\,,\\
 {R^4}_{i4i}&=\,\frac{2a\,[2(x^i)^2-(x^j)^2-(x^k)^2]}{r^3(r+a)^2}\,,&
{R^4}_{i4j}&=\ \,\frac{6a\,x^ix^j}{r^3(r+a)^2}\,.
\end{align}
The Ricci tensor vanishes.

\subsection{Momentum and spin}

In the above coordinate system, the (future pointing)  4-momentum of the photon is written as
\begin{equation}
P=\left(
\begin{array}{c}
\displaystyle
\frac{\bp}{B}\\[10pt]
\displaystyle
 \frac{\np}{A}
\end{array}\right)
\label{Pbis}
\end{equation}
with $\bp\in\bbR^3\setminus\{0\}$, the spatial linear momentum, and $\np\dpp=\sqrt{\bp\cdot\bp}$ (Euclidean scalar product). We suppose positive energy, $\np>0$. The 4-momentum is light-like, $P^2=0$.

The map $S$ is skewsymmetric with respect to the metric: $\rg(SV,W)=-\rg(V,SW)$ for all vectors $V$ and $W$. Accordingly, the spin tensor is defined by the Tulczyjew constraint SP=0.

For given $P$, the general solution of the Tulczyjew constraint is parametrized by the three components of the spin vector $\bs \in \bbR^3$ that we suppose non vanishing,
\begin{equation}
S=({S^\mu }_\nu)=\left(
\begin{array}{cc}
j(\bs)&\displaystyle
-\frac{(\bs\times\bp)}{\np}\frac{A}{B}\\[6pt]
\displaystyle
-\frac{(\bs\times\bp)^T}{\np}\frac{B}{A}&0
\end{array}\right).
\label{Sbis}
\end{equation}
The vector-product is with respect to the Euclidean metric and we define the linear map $j(\bs):\bp\mapsto\bs\times\bp$. We have,
\begin{equation}
-\half\Tr(S^2)=s^2\,,
\end{equation}
with the longitudinal spin, or ``scalar'' spin,
\begin{equation}
s\dpp=\frac{\bs\cdot\bp}{\np},
\label{scalarspin}
\end{equation}
which turns out to be a constant of the system \cite{Sou74}. The scalar spin $s$ is not to be confused with the norm $\Vert\bs\Vert$ of the spin vector. The helicity or handedness of the photon is $\sign(s)$.

%

In the Schwarzschild metric we obtain with the notations (\ref{notations}),
\begin{align}
\Pf(R(S))&=\,\frac{48\, a^2r^4}{(r+a)^{12}\np}\lb\bx \times\bp\cdot\bs\rb(\bs\cdot\bx),\\[2mm]
R(S)(S)&=\,\frac{8\, ar}{(r+a)^6}\, \lb 3\lb\bx \times\bp\cdot\bs\rb^2/\np^2-3 (\bs\cdot\bx)^2+s^2r^2\rb,\\[2mm]
S\,R(S)\,P&=\dpp
\begin{pmatrix}{\bf c}\\d
\end{pmatrix}\qq{\rm with} \\[2mm]
\nonumber
{\bf c}&=\,\frac{12\,a\,r^3}{(r+a)^8}\,\Bigg[(\bs\cdot\bx)^2{\bp}\,-\np s\,(\bs\cdot\bx )\,\bx-\lb\bx \times\bp\cdot\bs\rb\,\bs\times\bx\\[2mm]
&\qq\qq\qq\qq\qq\qq\qq\qq\qq\qq\qq\qq\qq+\lb\bx \times\bp\cdot\bs\rb\lp\bx\cdot\frac{\bp}{\np}\rp\bs \times\frac{\bp}{\np}\Bigg]\,,\\[2mm]
d&=\,\frac{12\,a\,r}{(r-a)(r+a)^5}\,\Big[ s\,(\bs\cdot\bx)\lp\bx\cdot{\bp}\rp-\ns^2/\np\,\lp\bx\cdot{\bp}\rp^2
\nonumber
\\[2mm]
&\qq\qq\qq\qq\qq\qq\qq\qq\qq
+\np\,(\ns^2-s^2)\,r^2-2\lb\bx \times\bp\cdot\bs\rb^2/\np\Big] .
\end{align}
The following vector identity will be useful:
\bb
\lb\bu \times\bv\cdot\bw\rb^2&=&\Vert\bu\Vert^2\Vert\bv\Vert^2\Vert\bw\Vert^2
+
2\,(\bu\cdot\bv)(\bu\cdot\bw)(\bv\cdot\bw)\nonumber\\
&&-\Vert\bu\Vert^2(\bv\cdot\bw)^2
-\Vert\bv\Vert^2(\bu\cdot\bw)^2
-\Vert\bw\Vert^2(\bu\cdot\bv)^2\,.
\ee

\subsection{Conservation laws}

The group of isometries of Schwarzschild spacetime is $\rO(3)\times\bbR$, its generators are the Killing vector fields of the metric~(\ref{gbis}),
$Z=\varepsilon^i_{\,jk}\,\omega^jx^k\,\partial/\partial x^i+\epsilon \,\partial/\partial t $,
where $\bomega\in\bbR^3$ and $\epsilon \in\bbR$ stand for infinitesimal rotations and time translations, respectively; the $\varepsilon^i_{\,jk}$ are the structure constants of $\so(3)$.
Using the general expression \cite{Sou70}
\begin{equation}
\Psi(Z)=P_\mu{}Z^\mu+\half{}S^{\mu\nu}\nabla_\mu{}Z_\nu
\label{Psi}
\end{equation}
 of the ``moment map'', $\Psi$, associated with a Killing vector field, $Z$, together with the expressions (\ref{Pbis}) and (\ref{Sbis}) for $P$ and $S$, we find in a straightforward fashion
$
\Psi(Z)=-\bcL\cdot\bomega+\cE\,\epsilon
$
where 
\begin{equation}
\cE=
\,\frac{r-a}{r+a}\,\np+\,\frac{2ar}{(r+a)^4\np}\lb\bx \times\bp\cdot\bs\rb\,,  
\label{cE}
\end{equation}
is the conserved energy and
\begin{equation}
\bcL=\lp\frac{r+a}{r}\rp^2 \bx\times\bp+\,\frac{r-a}{r+a}\, \bs\,+\,\frac{2a}{r^2(r+a)}\,(\bs\cdot\bx)\, \bx,
\label{cL}
\end{equation}
 the conserved angular momentum featuring both an extra spin contribution. The latter equation defines an affine map between spin and angular momentum. We will use its inverse:
 \bb
 \bs\,=\,\frac{r+a}{r-a}\, \lb \bcL-\lp\frac{r+a}{r}\rp^2\,\bx\times\bp-\,\frac{2a}{r^2(r+a)}\, (\bcL\cdot\bx)\,\bx\rb. \label{bs}
 \ee 

\subsection{Specifying the Souriau-Saturnini equations}

Now that we have defined the metric together with the objects appearing in the equations of motion and the conserved quantities, we are ready to spell out the Souriau-Saturnini equations (\ref{dotXter} - \ref{dotSter}) for the case of Schwarzschild spacetime.

Let us introduce the shorthand,
\begin{equation}
\label{def_d}
D\dpp=r^2(\bs\cdot\bp)-3(\bp\cdot\bx)(\bs\cdot\bx).
\end{equation}

To obtain the equations of motion in $3$-space, we trade the affine parameter $\tau$ for the coordinate time $t$ using (\ref{dotXter}),

\begin{equation}
\label{dtdtau}
\frac{dt}{d\tau}= \frac{r+a}{r-a}\,\np
\!\left[
\frac{s\,D \, \np}{s^2r^2\np^2-3(\bs\cdot\bx)^2\np^2+3[\bx\times\bp\cdot\bs]^2}
\right]\,,
\end{equation}
which we assume non-vanishing. 
By abuse of notation we write $\tau(t)$ for the inverse function of~$t(\tau)$ and we do not distinguish $\bx=\bx(t)=\bx(\tau(t))$ and likewise for $\bp$ and $\bs$. Then we have, from the Souriau-Saturnini equations (\ref{dotXter} - \ref{dotSter}) and (\ref{dtdtau}):
\begin{align}
\,\frac{d\bx}{dt}\,
& =\,\frac{r^2(r-a)}{\np\,(r+a)^3D}
&\hspace{-10mm}\Big\{&
r^2(\bs\cdot\bp)\,\bp-3\np^2(\bs\cdot\bx)\,\bx+3[\bx\times\bp\cdot\bs]\,\bx\times\bp\Big\}\,, \label{dx_spin}\\
\,
\frac{d\bp}{dt}\,
& =\,\frac{2\,a}{\np\,(r+a)^4D}
&\hspace{-10mm}\Bigg\{&
r^2(r-a)\left[(\bs\cdot\bp)(\bp\cdot\bx)-\frac{3r}{(r+a)^3}\,(\bs\cdot\bx)[\bx\times\bp\cdot\bs]\right]\,\bp  \qq\qq\nonumber\\
&&&\qq\qq\qq\qq-r\,\np^2\big[D+r\,(r-a)\,(\bs\cdot\bp)\big]\,\bx  \nonumber\\
&&&\qq\qq\qq\qq\qq\qq+3\,(r-a)[\bx\times\bp\cdot\bs]\,(\bp\cdot\bx)\,\bx\times\bp\Bigg\},\label{dp_spin}\\
\,\frac{d\bs}{dt}\,
& =\,\frac{1}{\np\,(r+a)^4D}
&\hspace{-10mm}\Big\{&
3(r-a)(r+a)^3\big[\lp-r^2 \np^2+(\bx\cdot\bp)^2\rp \bs\times\bp\nonumber\\
&&&\hspace{-10mm}+\lp 2 \np^2(\bx\cdot\bs)-(\bx\cdot\bp)(\bs\cdot\bp)\rp \bx\times\bp\big]+ 2 a r D\lp(\bx\cdot\bs)\bp-(\bx\cdot\bp)\bs\rp\nonumber\\
&&&+ 2 a(r-a)\big(-r^2 (\bs\cdot\bp)^2\bx-3[\bx\times\bp\cdot\bs]^2 \bx\nonumber\\
&&&\qq\qq+r^2(\bx\cdot\bs)(\bs\cdot\bp) \bp+3[\bx\times\bp\cdot\bs](\bx\cdot\bs) \bx\times\bp\big)\Big\}.\label{ds_spin}
\end{align}

With the equations above, we can verify that the conserved quantities, namely the scalar spin (\ref{scalarspin}), the energy (\ref{cE}) and the total angular momentum (\ref{cL}) are conserved. We have indeed $d\cE/dt = d\bcL/dt = ds/dt = 0$.

We can simplify the system by only considering the equations of position and momentum (\ref{dx_spin}, \ref{dp_spin}) and by eliminating $[\bx\times\bp\cdot\bs]$ and $(\bs\cdot\bx)$ in favour of  the conserved angular momentum $\bcL$ using equation (\ref{bs}) and by eliminating $(\bs\cdot\bp)$ in favour of the conserved scalar spin $s$ using equation (\ref{scalarspin}). We use the following relations
\bb
\bx\times\bp\cdot\bs &=& \frac{r+a}{r-a}\lp \bx\times\bp\cdot\bcL-\lp\frac{r+a}{r}\rp^2\lp r^2\np^2-(\bx\cdot\bp)^2\rp\rp
\label{xps}
,\\[6pt]
\bs\cdot\bx&=&\bcL\cdot\bx,\\[6pt]
\bs\cdot\bp&=&s\np. \label{sp}
\ee
We are thus left with six equations for six unknown functions of $t$, which will be spelled out later, (\ref{pert_dx}, \ref{pert_dp}).

We also have a formula for the norm of $\bp$ from the conserved quantities (\ref{cE}) and (\ref{cL}),
\begin{equation}
\np = \frac{r-a}{r+a} \; \frac{(r+a)^2 \cE - \frac{2 a r}{(r^2-a^2) \np} \lp \bx\times\bp\cdot\bcL\rp}{(r-a)^2 \; - \; \frac{2a}{r \np^2} \Vert\bx\times\bp\Vert^2}
\label{redshift}
\end{equation}
and 
\begin{align}
\frac{d}{dt}\lp \frac{\bp}{\np} \rp
= \, \frac{2 a}{(r+a)^4 D} \Bigg\{ &\Big(3r(\bs\cdot\bx) (\bx\cdot\bp)-(2r-a) (\bs\cdot\bp)r^2\Big)  \left(\bx-\frac{(\bx\cdot\bp)\, \bp}{\np^2}\right) \nonumber \\ 
& \qquad +3(r-a) \frac{(\bx\cdot\bp)}{\np^2} \lb\bx\times\bp\cdot\bs\rb \, \bx\times\bp \Bigg\}.
\end{align}
Noticing that this last equation and the three equations for position only depend on $\bp/\np$ our system effectively reduces to five equations.

The results above can already by found in Saturnini's thesis \cite{Sat76} of 1976.

 
\subsection{Radial case}
The first observation is that in the radial case, \textit{i.e.} with an initial momentum parallel to the initial position, the equations of motion (\ref{dx_spin} - \ref{ds_spin}) reduce to those of the radial geodesics,
\begin{align}
\frac{d\bx}{dt} & = \frac{r^2(r-a)}{(r+a)^3} \frac{\bp}{\np}, \label{radial_dx} \\
\frac{d\bp}{dt} & = - \frac{2\,a\, r^2}{(r+a)^4} \, \bp, \label{radial_dp} \\
\frac{d\bs^\perp}{dt} & = - \frac{2\,a\, r^2}{(r+a)^4} \, \bst \label{radial_ds}.
\end{align}
While the differential equation (\ref{radial_dp}) displays the well known redshift effect of light, it is striking that we have the same expression (\ref{radial_ds}) for the evolution of the transverse spin. This can be expected when looking at the Souriau--Saturnini equations (\ref{dotXter} - \ref{dotSter}) and noticing that the redshift terms in (\ref{radial_dp}) and (\ref{radial_ds}) come from the covariant derivative. Indeed, when the photon is following the geodesic trajectory, the Souriau-Saturnini equations reduce to the geodesic equations \textit{i.e.} $\dX = P$ and $\dP = \dS = 0$, meaning that both $P$ and $S$ are parallel transported.

We also take the opportunity to note that equation (\ref{cE}) tells us that the conserved energy $\cE$ is modified by the transverse spin in general, but not in the radial case.
 
\section{Null geodesics \& spinless gravitational lensing}
\label{s:3}

In this section, we first show that we recover the known spinless, massless case, albeit in a slightly different form than the usual geodesic equations, by putting $\bs = 0$ in our equations, and we rederive the well known deviation angle $\Delta \varphi$.

\subsection{Some preliminaries}

If we put $\bs=0$ in (\ref{cE}) and (\ref{cL}), the Noether quantities are of the form
\begin{equation}
\cE=
\frac{r-a}{r+a}\,\np\,
\qquad
{\rm and}
\qquad  
\bcL=\lp\frac{r+a}{r}\rp^2 \bx\times\bp\,.
\label{cEcLs=0}
\end{equation}
From $dX/d\tau=P$, and equation (\ref{Pbis}), we find
\bb
\bp=\left(\frac{r+a}{r}\right)^2\frac{d\bx}{d\tau}\,.
\label{bps=0}
\ee
For null geodesics, $P^2=0$, we have
\bb
\left\Vert\frac{d\bx}{d\tau}\right\Vert=\frac{\cE\,r^2}{r^2-a^2}\,.
\label{normdxdtaus=0}
\ee
Taking advantage of the conservation of total angular momentum, $\bcL$, we compute $\bx\times\bcL$ and end up with
\bb
\frac{d\bx}{d\tau}=-\frac{r^2}{(r+a)^4}\,\bx\times\bcL+\lambda\,\bx
\label{dxdtau}
\ee
where the function $\lambda$ satisfies (using (\ref{bps=0}),  (\ref{dxdtau}) and (\ref{normdxdtaus=0}))
\bb
\lambda =\frac{\bx\cdot\bp}{(r+a)^2}\qq\text{and}\qq
\lambda^2=\frac{\cE^2\,r^2}{(r^2-a^2)^2}-\frac{\cL^2\,r^4}{(r+a)^8}
\ee
with $\cL=\Vert\bcL\Vert$. We note that $\lambda^2\ge0$ implies a condition on $\cE,\cL$ and $r$. By taking the scalar product on both sides of equation (\ref{dxdtau}) with $\bx$, we obtain the simple expression
\bb
\frac{dr}{d\tau}=\lambda\,r\,.
\label{drdtau}
\ee
We record for further use that
\bb
\frac{d\lambda}{d\tau}=-r^2\left[\frac{\cE^2(r^2+a^2)}{(r^2-a^2)^3}-\frac{2\cL^2r^2(r-a)}{(r+a)^9}\right].
\label{dlambdadtau}
\ee
Let us stress that the latter equations lead precisely to the equations of null geodesics given in terms of the Christoffel symbols (\ref{Gamma}). Here we used, instead, the conservation laws, including a number of computational tricks, to obtain the velocity (\ref{dxdtau}). Note that the time-component of the geodesic equation yields (up to a global sign):
\bb
\frac{dt}{d\tau}=\cE\left(\frac{r+a}{r-a}\right)^2
\label{dtdtaus=0}
\ee
which is clearly non-vanishing. Comparison with the general equation (\ref{dtdtau}), which is ill-defined in the limit $s\to0$, shows a striking similarity with equation (\ref{dtdtaus=0}), namely the latter  is identical to the former provided we ignore the spin-dependent factor on the RHS.

To make the link with equations (\ref{dx_spin}) and (\ref{dp_spin}), let us write down the equations of motion of the null geodesic in the form:
\begin{align}
\,\frac{d\bx}{dt}\,
& =\,\frac{r^2(r-a)}{(r+a)^3\np}\,\bp\,,\label{dxdtgeo}
\\[3mm]
\,\frac{d\bp}{dt}\,
& =\,\frac{2a}{(r+a)^4\np}
\Big\{(r-a)(\bp\cdot\bx)\,\bp \, -\,(2r-a)\,\np^2\,\bx  
\Big\}.\label{dpdtgeo}
\end{align}

\subsection{Lensing in weak fields}
\label{sec_deltaphi}
We restrict our analysis to geodesics remaining in regions of space where the gravitational field is weak, \textit{i.e.} where all distances $r(t)$ remain much larger than the Schwarzschild radius $a$,  
\bb \alpha (t)\dpp=\,\frac{a}{r(t)}\,  \ll\,1,\ee
and linearize with respect to $\alpha $. We take our initial conditions at $\tau=t=0$:
\bb \bx_0=
\begin{pmatrix}
-x_0\\b\\0
\end{pmatrix} \qq{\rm and}\qq
\bp_0=
\begin{pmatrix}
 p_0\\0\\0
\end{pmatrix}.
\ee
To alleviate notations we will write from now on $\bx=(x_1,\,x_2,\,x_3)$ with lower indices.
Following tradition we consider the photon in the $x_1$-$x_2$ plane with energy $p_0>0$
 coming in from the left, $x_0>0$, with positive impact parameter $b$. We suppose $a\ll b$. Then we have to first order in $\alpha $:
 \bb \cE\sim(1-2\alpha _0) p_0,\qq \,\frac{\bcL}{\cE}\, \sim\,-(1+4\alpha _0)\,b\begin{pmatrix}
0\\0\\1
\end{pmatrix} \ {\rm and}\ \lambda \sim \pm\,\frac{\cE}{r}\,  \sqrt{1-\lp\frac{\cL}{\cE}\rp^2\frac{1-8\alpha }{r^2}} . 
\ee
The equations of motion (\ref{dxdtau}) become:
\begin{align}
\dot x_1&\sim-\cL(1-4\alpha )\,\frac{x_2}{r^2}\, +\lambda x_1,\\
\dot x_2&\sim+\cL(1-4\alpha )\,\frac{x_1}{r^2}\, +\lambda x_2,\\
\dot x_3&=0,
\end{align} 
implying
\bb
\dot\varphi=\,\frac{x_1\dot x_2-x_2\dot x_1}{r^2}\,  \sim -\,\frac{1-4\alpha }{r^2}\,  \cL
\ee
Equation (\ref{drdtau}), $\dot r =\lambda r$, tells us that the distance of closest approach $r_p$ (`perihelion')  is reached when $\lambda $ vanishes. Therefore 
\bb\cL/\cE\sim b\sim (1+4\,\alpha _p)\,r_p,\ee
and $r_p\sim b-4\,a+4\,ab/r_0$.

Our aim is to compute the scattering angle $\Delta \varphi $ for $x_0\rightarrow\infty$. As we have set the cosmological constant to zero, spacetime is flat far away from the mass and there coordinate and physical angles coincide. Denoting by $\varphi _p$ the angle of closest approach, we have $\Delta \varphi =\pi -2\varphi _p$. We can compute $\varphi _p$ by integrating
\begin{align}
 \,\frac{d\varphi }{dr}\,& =\,\frac{\dot \varphi }{\dot r}\,=\,\frac{\dot\varphi }{\lambda r}\,
\sim\mp
\,\frac{1-4\alpha }{r^2}\,\frac{\cL}{\cE}\,\lb1-\lp\frac{\cL}{\cE}\rp^2\frac{1-8\alpha }{r^2}\rb^{-1/2}\nonumber \\[1mm]
&\sim\,\mp \,\frac{1+4\alpha _p}{r_p}\,\frac{1-4\alpha }{r/r_p}\lb(r/r_p)^2-1\rb^{-1/2}\lp1-4\,\frac{\alpha -\alpha _p}{(r/r_p)^2-1}\rp \label{dphidr}
\end{align} 
between $r_0=\infty$ and $r_p$.   
In this interval both $r$ and $\varphi $ decrease and we must choose the positive signs in equation (\ref{dphidr}). Our initial angle is $\varphi _0=\pi $ and we obtain with $u\dpp=r/r_p$,
\bb
\pi -\varphi _p\sim (1+4\alpha _p)\int_{1}^{\infty}
\frac{1-4\alpha_p/u }{u}\lb u^2-1\rb^{-1/2}\lp1-4\alpha_p\,\frac{1/u -1}{u^2-1}\rp du\,=\,
\,\frac{\pi }{2}\, +4\,\frac{a}{r_p}\,. 
\ee
Note the integrable singularity at the perihelion, $u=1$.
Finally, in linear approximation, the scattering angle takes its famous value: $\Delta \varphi \sim 4\,GM/r_p$.

We thus recover the known geodesic equations in the Schwarzschild metric, and the well known deflecting angle $\Delta \varphi$, from the Souriau-Saturnini formalism and putting $\bs = 0$. The resulting equations of motions (\ref{dxdtgeo} - \ref{dpdtgeo}) are first order equations, but are strictly equivalent to the second order geodesic equations. Now, the next step is to consider the spinning case, $\bs \neq 0$, hence considering the full equations of motions (\ref{dx_spin} - \ref{ds_spin}). This is done in the following sections.

\section{Numerical solutions}
\label{s:4}

Since solving the system of equations (\ref{dx_spin}, \ref{dp_spin}) is not straightforward, we will use the help of numerical integration to propagate specific initial conditions. These numerical solutions will guide us towards perturbative ones.

The numerical integration meets the usual problem of
 accuracy errors when computing the difference of two almost identical numbers. It becomes relevant here because the present system of equations involves such computations, especially when conserved quantities are involved, \textit{e.g.} (\ref{xps}). This is why it is better to numerically solve all of the 9 differential equations (\ref{dx_spin} - \ref{ds_spin}), including those of the spin.

Even with such measures, integrating these equations over a long time can be tricky with Mathematica. The step algorithm seems overly cautious and is eager to stop the integration process due to stiffness problems, even though all quantities involved are well defined, finite, and smoothly evolving. We need to select the right precision parameters to keep the step algorithm from stopping the integration. Yet, this does not create instabilities in the trajectory of the simulation and we obtain very precise results.

It is convenient to take the initial conditions not at infinity but at perihelion $r_0=r_p$ of the trajectory of the photon around the star located at the origin:
\begin{equation}
\bx_0 = \lp
\begin{array}{c}
r_0\\
0\\
0
\end{array}\rp,\qquad
\bp_0 = \lp
\begin{array}{c}
0\\
p_0\\
0
\end{array}\rp,\qquad
\bs_0 = \lp
\begin{array}{c}
0\\
s\\
s^\perp_0
\end{array}\rp.
\label{ic}
\end{equation}
Note that the first component of the initial transverse spin $\bs^\perp_0$ vanishes, because at perihelion ${d\bx}/{dt}|_0\cdot\bx_0 = 0$.

We use SI units here. The photon starts with a wavelength of $\lambda_0 = 600\,$nm and a helicity of $\chi = +1$, the star has a Schwarzschild radius of $a = 3\cdot10^3\,$m, and the initial distance from the center of the star to the perihelion is~$r_0 = 3\cdot10^5\,$m. The numerical integration runs from $0$ to $0.1\,$s. While we have $s = \hbar$ in the initial conditions (\ref{ic}), we will put $s_0^\perp = 0$ for the time being, because otherwise the trajectory leaves the neighbourhood of the geodesic when $s_0^\perp$ is close to $\hbar$. We will come back to the transverse spin in the perturbative analysis in the next section.

\begin{figure}
\centering
\includegraphics[scale=0.9]{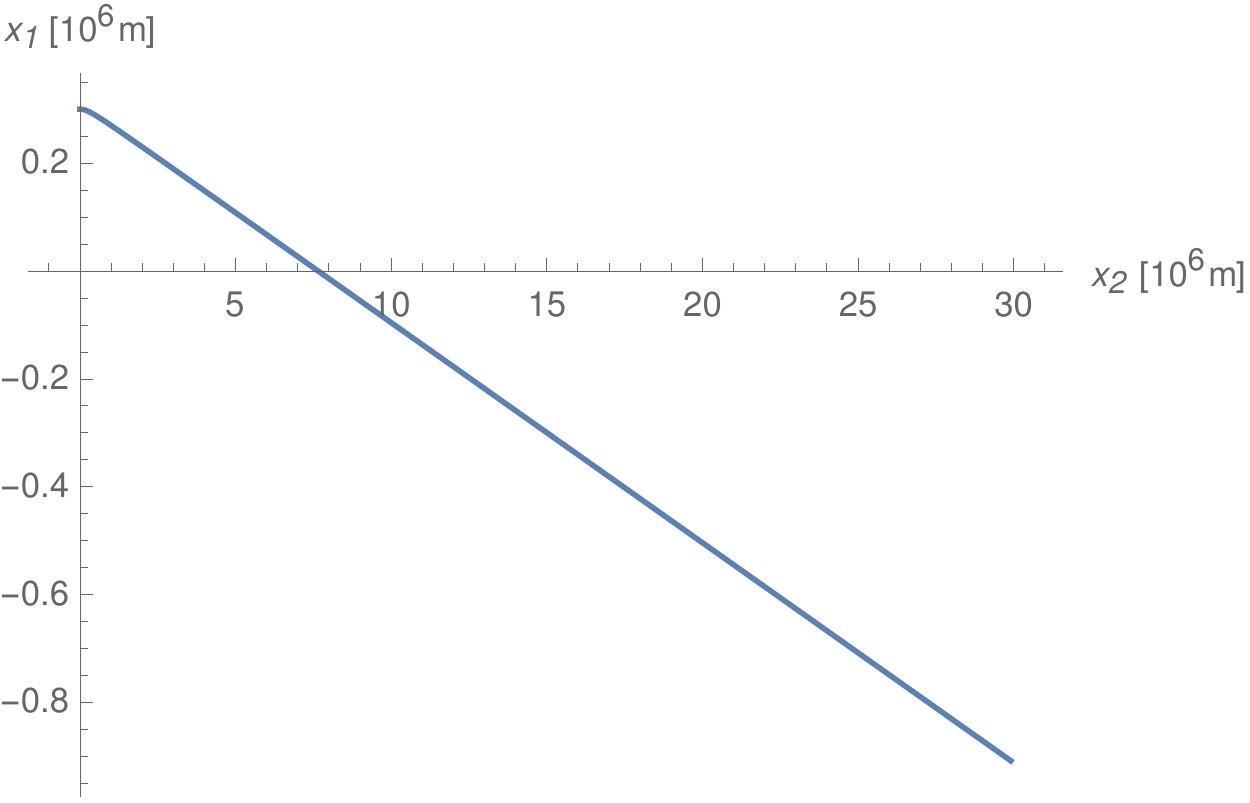}
\caption{Trajectory of the spinning photon in the geodesic plane. Visually, this trajectory is the same as that of the spinless photon.}
\label{fig_geod}
\end{figure}

\begin{figure}
\centering
\includegraphics[scale=0.9]{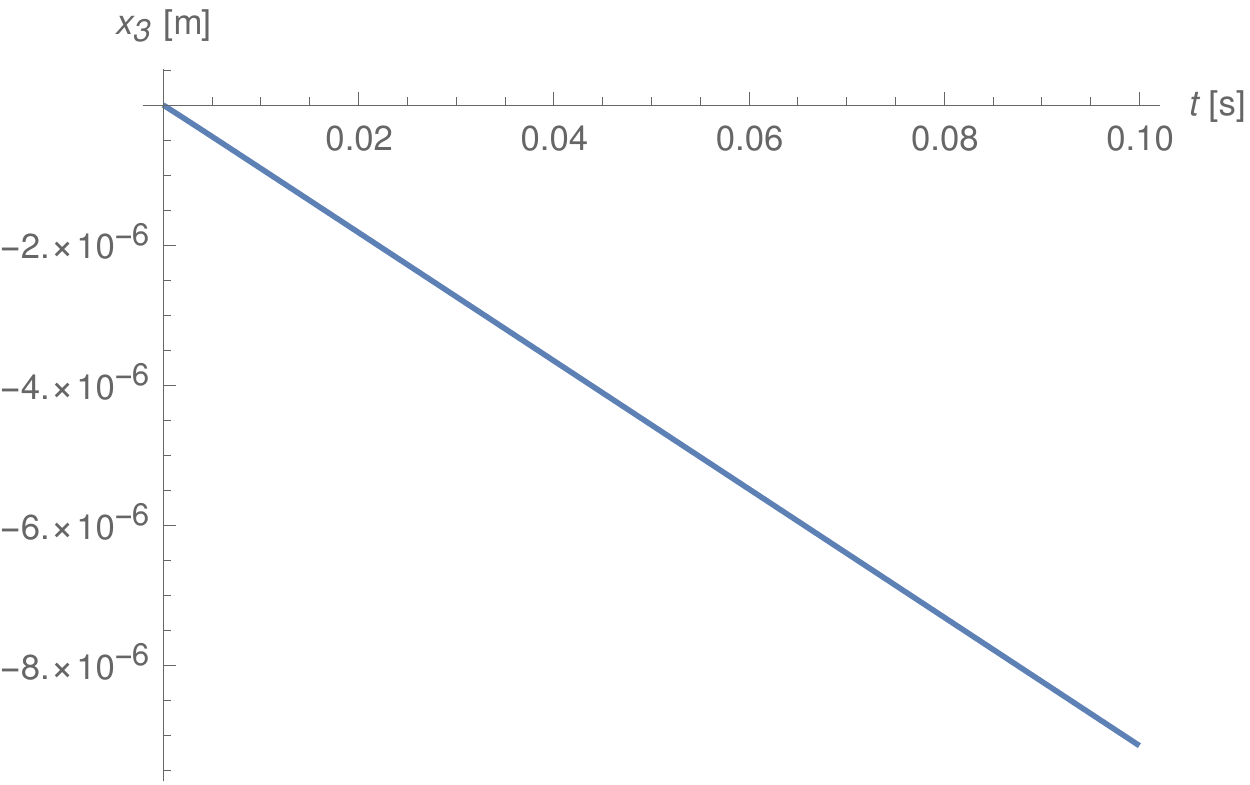}
\caption{Component $x_3$ of the trajectory of the photon as a function of time. The spinning photon leaves the geodesic plane, albeit with a very small angle.}
\label{fig_x3t}
\end{figure}

\begin{figure}
\centering
\includegraphics[scale=0.9]{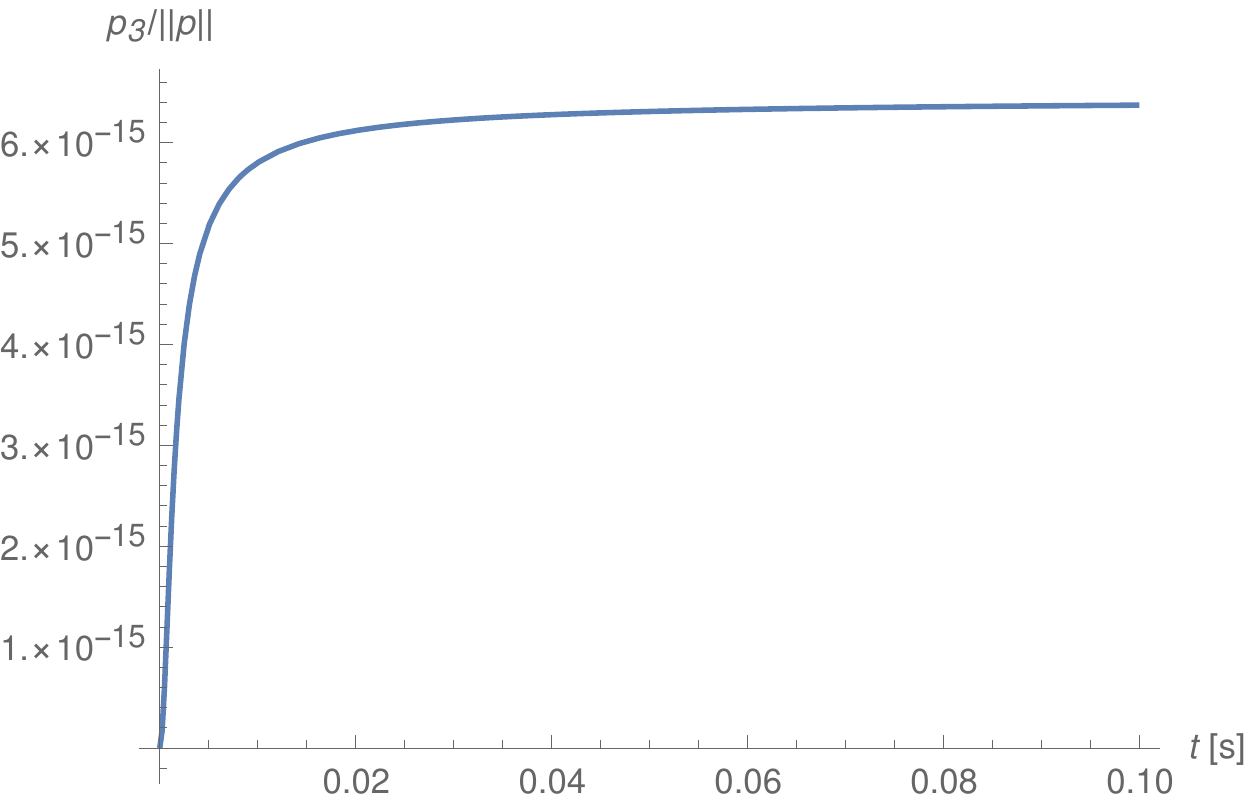}
\caption{Component $p_3$ of the momentum of the photon, normalized to the norm of the momentum, as a function of time. Just like with $x_3$, while the geodesic momentum is contained in the plane $(p_1, p_2)$, the momentum of the spinning photon has a component perpendicular to that plane. Notice that the sign of $p_3$ is opposite to that of $x_3$.}
\label{fig_p3t}
\end{figure}

Figure \ref{fig_geod} shows the trajectory of the spinning photon in the geodesic plane. This trajectory is almost identical to the null geodesic one. Indeed, the difference between the coordinates $x_1$ and $x_2$ of spinning and spinless photons is of the order of the nanometer at the end of the numerical integration. The main differences are the transverse components $x_3$ of the trajectory, and $p_3$ of the momentum, pictured in Figures \ref{fig_x3t} and \ref{fig_p3t} respectively. While the geodesic trajectory is contained within the plane $(x_1, x_2)$, the equations of motion (\ref{dx_spin} - \ref{ds_spin}) imply non-vanishing transverse components $x_3$ and $p_3$.

The angle $\beta$ of the trajectory going out of the plane is small, but constant. As shown in figure \ref{fig_x3t}, it is about $\beta = - 6.3 \cdot 10^{-8}\, \arcsec$. The sign of the angle $\beta$ depends directly on the helicity $\chi$. Indeed when changing the helicity from $+1$ to $-1$, the amplitude of the angle stays the same, but the sign switches. We see from numerical integrations that the trajectories of two different helicity photons are  symmetric with respect to the null geodesic. The transverse momentum $p_3$ also shows the same behavior under helicity changes and its sign is again opposite to that of $x_3$.

In the next section, we will confirm and explain these results with a perturbative approach.

\section{Perturbative solutions}
\label{s:5}

We wish to compare the behavior of our system (\ref{dx_spin}, \ref{dp_spin}) describing the trajectories of photons with their due spin to the behavior of null geodesics.

Now, define two {\it constant} small parameters,
\begin{equation}
\label{params}
\alpha = \frac{a}{r_0} \qquad {\rm and} \qquad \epsilon = \frac{\hbar}{r_0\, p_0},
\end{equation}
where, for the sun, $\alpha$ is typically of the order of $10^{-6}$ and $\epsilon$ of the order of $10^{-16}$ for photons in the visible spectrum. A small $\epsilon$ corresponds to photons having a  wavelength much smaller than its distance to the star, which is a sensible hypothesis. Due to the particularities of this system of equations, namely $D$ (\ref{def_d}) being of order $\epsilon$, we must consider second order terms in $\epsilon$ to obtain the first order equations.  In $\alpha$, linear terms will be sufficient.

Let us redefine the spin by setting
\begin{equation}
s =\dpp \chi \hbar\qquad {\rm and} \qquad s^\perp_0=\dpp w\hbar,
\end{equation}
where $\chi = \pm 1$ is the helicity of the photon and $w$ is finite and dimensionless. 
We easily obtain the conserved quantities (\ref{cE}) and (\ref{cL}) from  the initial conditions (\ref{ic}),
\begin{equation}
\cE \sim (1-2 \alpha) \, p_0 \qquad  {\rm and}  \qquad \bcL \sim r_0\,p_0 \lp
\begin{array}{c}
0\\
(1-2\alpha) \, \chi\,\epsilon\\
(1+2\alpha)+(1-2 \alpha )\,w\,\epsilon
\end{array}\rp
\label{pert_qtt}
\end{equation}
We define the normalized quantities,
\begin{equation}
  \xl = \frac{\bx \cdot \bcL}{r\, \chi s}
  \qquad  {\rm and}  \qquad
  \xps = \frac{\bx \times \bp \cdot \bs}{r\, p\,\chi s}.
\end{equation}
We can then write the equations (\ref{dx_spin} - \ref{dp_spin}) as
\begin{align}
\frac{d\bx}{dt}\,& =\,
\frac{r^2(r-a)}{(r+a)^3\lp r\,\np-3(\bx\cdot\bp)\,\xl\rp} \lb r\, \bp-3\np \, \xl \,\bx+3\,\xps\, \bx\times\bp\rb \label{pert_dx},\\
\frac{d\bp}{dt}\,& =\,
\frac{2\,a}{(r+a)^4\lp r\,\np-3(\bx\cdot\bp)\,\xl\rp} \Big[ r(r-a)\lp(\bx\cdot\bp)-3 \,\frac{r^3}{(r+a)^3}\,  s \,\chi \, \xl \, \xps \rp\bp \nonumber \\
& \hspace{6.5cm} -r \np\lp(2r-a) \np-3(\bx\cdot\bp) \, \xl\rp\,\bx \nonumber \\
& \hspace{6.5cm} + 3 (r-a) \, \xps \, (\bx\cdot\bp) \,\bx\times\bp \, \Big]\label{pert_dp}.
\end{align}
Let us momentarily forget the physical aspect of this system and set $\epsilon = 0$ in (\ref{pert_qtt}). Then with the initial conditions (\ref{ic}) the differential equations (\ref{pert_dx}) and (\ref{pert_dp})  reduce to those of the null geodesics (\ref{dxdtgeo}) and (\ref{dpdtgeo}). Indeed, from (\ref{xps}), we have initially $\xps|_0 = 0$ and $\xl|_0 = 0$, reducing the initial system to the geodesic one. If we are on a geodesic trajectory, which is in the plane spanned by $\bx_0$ and $\bp_0$, then  $\xl=0$ and $\xps = 0$ continue to vanish due to geodesic conservation of angular momentum and the photon continues on the geodesic trajectory.

This heuristic argument and our numerical results in the last section motivate the ansatz
\begin{equation}
\bx \sim \lp\begin{array}{ccccc}
x_1 & + & \epsilon\, y_1 & + & \epsilon^2\, z_1\\
x_2 & + & \epsilon\, y_2 & + & \epsilon^2\, z_2\\
& & \epsilon\, y_3 & + & \epsilon^2\, z_3
\end{array}\rp
\quad {\rm and} \quad
\bp \sim \lp\begin{array}{ccccc}
p_1 & + & \epsilon\, q_1 & + & \epsilon^2\, u_1\\
p_2 & + & \epsilon\, q_2 & + & \epsilon^2\, u_2\\
& & \epsilon\, q_3 & + & \epsilon^2\, u_3
\end{array}\rp\,,
\label{ansatz}
\end{equation}
where $x_1,\, x_2,\, p_1,\, p_2$ solve the geodesic equations. Define $r_g = \sqrt{x_1^2+x_2^2}$ and similarly for $p_g$.
To leading order, we have
\begin{align}
\xl & = (1-2 \alpha) \, \frac{x_2}{r_g}+(1+2\alpha) \, \chi \frac{y_3}{r_g} + \cO(\epsilon), \label{pert_lx}\\
\xps & = \,\chi \,\frac{r_0p_0}{r_gp_g}\, \Bigg[
w+2\,\frac{a}{r_g}\,\frac{x_1 y_1 + x_2 y_2}{r_g^2}\, -\lp1 + 2\alpha + 2\frac{a}{r_g}\rp\,\frac{y_1p_2-y_2p_1+ x_1q_2-x_2q_1}{r_0p_0}\Bigg]
+ \cO(\epsilon). \label{pert_xps}
\end{align}

In order to recover the geodesics in the limit $\epsilon \rightarrow 0$, we thus need these two leading terms to be zero implying the initial transverse spin to vanish and some conditions on first order terms in $\epsilon$ that are valid at least to first order in $\alpha $:
\begin{align}
w&\sim 0,\label{w}\\
y_3 & \sim -\chi \,(1-4\alpha) x_2, \label{cond_y3}\\
x_1 \, y_1  + x_2 \, y_2& \sim 0,\\
y_1p_2-y_2p_1+ x_1q_2-x_2q_1 & \sim 0.
\label{constraints}
\end{align}

Plugging the ansatz (\ref{ansatz}) into the six scalar equations (\ref{pert_dx}) and (\ref{pert_dp}) we obtain twelve equations: six in $\epsilon^0$ and six in $\epsilon^1$. The six equations in $\epsilon^0$ are equivalent to the four equations (\ref{w} - \ref{constraints}). The six equations in $\epsilon^1$ yield:
\bb y_1\sim y_2\sim q_1\sim q_2\sim { q_3\sim  \cO(\alpha )\qq\qq{\rm and}\qq\qq z_3\sim \cO(\alpha )}.\ee
At this point, we may even obtain the terms of order $\alpha \epsilon$ giving us constraints on $z_1$ and on the initial transverse spin and we end up with  $y_1 \sim y_2\sim q_1\sim q_2\sim \cO(\alpha^2)$ and 
\begin{align}
\epsilon \, y_3 & = - \epsilon \, \chi\lp(1-4\alpha)\,t-4 \, \alpha \, r_0 \, \ln\frac{t+\sqrt{r_0^2+t^2}}{r_0}\rp, \\
\epsilon \, q_3 & = 2 \, \epsilon \, \alpha \, \chi \, p_0 \left(1-\frac{r_0}{\sqrt{r_0^2+t^2}}\right),
\end{align}
and our perturbative solution reads
\begin{align}
\bx & = \lp\begin{array}{c}
r_0 + 4 \, \alpha \, r_0\, \left(1-\frac{\sqrt{r_0^2+t^2}}{r_0}\right) \\[4pt]
t - 4 \, \alpha \, r_0 \ln\frac{t+\sqrt{r_0^2+t^2}}{r_0} \\[4pt]
- \epsilon \, \chi\lp(1-4\alpha)\,t-4 \, \alpha \, r_0 \, \ln\frac{t+\sqrt{r_0^2+t^2}}{r_0}\rp
\end{array}\rp + \cO(\epsilon^2, \alpha^2), \\[2mm]
\bp & = \lp\begin{array}{c}
- 4 \, \alpha \, p_0\frac{t}{\sqrt{r_0^2+t^2}} \\[4pt]
p_0 - 2 \, \alpha \, p_0 \lp 1 - \frac{r_0}{\sqrt{r_0^2 + t^2}}\rp \\[4pt]
2 \, \epsilon \, \alpha \, \chi \, p_0 \left(1-\frac{r_0}{\sqrt{r_0^2+t^2}}\right)
\end{array}\rp+ \cO(\epsilon^2, \alpha^2).
\label{pert_sol}
\end{align}
Finally, using (\ref{bs}) and $\bs = \frac{\bp}{\np} s + \bs^\perp$, we obtain the perturbative solution for the transverse spin,
\begin{equation}
\bs^\perp = \chi \, \hbar \lp\begin{array}{c}
- \frac{t}{r_0}(1-4 \alpha)+4 \alpha \ln \frac{t+\sqrt{r_0^2+t^2}}{r_0} \\[4pt]
- \frac{4 \alpha t^2}{r_0 \sqrt{r_0^2+t^2}} \\[4pt]
0
\end{array}\rp + \cO(\epsilon^2, \alpha^2).
\label{pertspin}
\end{equation}

The most striking effect of the spin on the trajectory of the photon is that it leaves the geodesic plane, but its projection on this plane coincides up to order $\epsilon \alpha$ with the geodesic. The angle $\beta $ between the  trajectory and the geodesic plane is obtained from $\beta \sim d(\epsilon y_3)/dx_2$ at infinity, which is immediate with the help of (\ref{cond_y3}),
\begin{equation}
\beta \sim - (1-4 \alpha) \frac{\chi \, \lambda_0}{2 \pi \, r_0}
\label{angle_beta}
\end{equation}
with the definition (\ref{params}) for $\epsilon$ and where $\lambda_0$ is the wavelength of the photon at perihelion. Notice that this angle depends both on the helicity of the photon $\chi = \pm 1$ and on its wavelength. Photons of the two different helicities  follow  symmetric trajectories with respect to the geodesic and the dependence on $\lambda_0$  produces a rainbow effect. In the case of the sun, with $r_0$ its radius, this means that two photons starting at the perihelion with opposite helicity will have an offset given by $2 \beta = 5.7 \cdot 10^{-11}\, \arcsec$. If these two photons then travel to the Earth, the offset between them would be of the order of $41 \mathrm{\mu m}$ in perfect conditions. The angle $\beta$ has the curious property of being independent of the mass of the star, at lowest order in $\alpha$. This seems to imply that this angle does not vanish as the mass of the star becomes arbitrarily small.  Let us note though, that the limit $\alpha \rightarrow 0$ is ill defined in the equations of motion and therefore in the perturbative solution. Indeed, the first of the Souriau-Saturnini equations (\ref{dotXter}) is independent of $a$ because both $R(S)(S)$ and $S R(S) P$ are proportional to $a$. The introduction of a cosmological constant will regularize this singularity, even at small scales, as will be shown in a forthcoming work.

Also, there is no correction of order $\epsilon \, \alpha$ to the usual deviation angle $\Delta \varphi$ in the plane, computed in subsection \ref{sec_deltaphi}.

Note that the transverse component of the momentum quickly reaches its maximum at a distance of a few $r_0$, which is $\epsilon \, {q_3}_{max} = 2 \, \epsilon \, \alpha \, \chi \, p_0$. Since the angle $\beta$ comes from a spin-orbit-like effect of the star on the trajectory, we would expect it to only act close to the star. To avoid this problem, we define $\gamma $ to be the angle between the geodesic plane and the momentum carried by the spinning photon. We have:
\begin{equation}
\gamma \sim \chi \frac{a \, \lambda_0}{\pi \, r_0^2}.
\label{angle_gamma}
\end{equation}
This angle does depend on the mass of the star and is even smaller than $\beta$. For the sun we have $2 \gamma = 4.9 \cdot 10^{-16}\, \arcsec$.

Our perturbative results for $y_3$ and $q_3$  above match our numerical results with a relative error of about $10^{-9}$ and $10^{-4.5}$, respectively. The match is better for $y_3$ because it contains terms of order 1 and of order $\alpha $, while $q_3$ is of order $\alpha$.

The analysis by Gosselin,  B\'erard \& Mohrbach \cite{gbm} starting from the Bargmann-Wigner equations shows birefringence with an angle equal to our $\gamma $ but in the geodesic plane.

\section{Conclusions}

For photons, quantum mechanics teaches us that the longitudinal component $s$ of the spin is $\pm \hbar$. This is in harmony with the conservation of $s$, which follows in general from the Souriau-Saturnini equations. Quantum mechanics also teaches us that the norm of the transverse spin $\nst$ is $ \hbar$. Two remarks arise from the present work. First, we saw in the radial case that the photon follows the null geodesic trajectory, and that the transverse spin undergoes the same evolution as the momentum: it is parallel transported. However, in our non-radial perturbative solution, equation (\ref{pertspin}), this norm vanishes at perihelion and then grows linearly with time~$t$ (to leading order). The linear growth implies that our perturbation theory breaks down for large times. This instability is absent from a generic Robertson-Walker metric where the norm of the transverse spin is proportional to the inverse Hubble parameter \cite{rw}.

With its continuously varying transverse spin, the instability reminds us of the instability of the classical hydrogen atom and its continuously varying energy. Indeed, the equations we use here are purely classical. The longitudinal spin is a constant of the system. Its definition comes from the co-adjoint representation of the Poincar\'e group \cite{Sou70,GQWood}. The transverse spin is not conserved. Its two degrees of freedom come from the introduction of the dipole moment. However, it is not clear from the geometrical derivation of the equations of motion, if these two degrees of freedom are the transverse spin in quantum mechanics. A way to determine their exact meaning would be to derive the Souriau-Saturnini equations (\ref{dotXter} - \ref{dotSter}) from quantum mechanics, \`a la eikonale. This is currently under investigation.

Notice also that the out-of-plane momentum is in the opposite direction with respect to the offset. This means that the star is intrinsicly acting on the photon's position and momentum, \textit{i.e.} a spin-orbit effect. Yet, at large time $t$ in the perturbative solution, we see that the trajectory's offset keeps increasing linearly, while the momentum stays constant and in the opposite direction. We would expect, once we are sufficiently far away from the star, that the star loses grip on the photon. Since spacetime is flat far away, we  expect the photon's momentum to carry the trajectory, which is not what we find. This is in line with the fact that we don't recover the equations of motion in flat spacetime in the limit $a \rightarrow 0$. We will see in a future work that including the cosmological constant helps us mitigating this problem, namely recovering null geodesics ``far away'' from the star, even at scales much smaller than those the cosmological constant would suggest. The cosmological constant also puts an upper bound to the transverse spin, whose value still remains in conflict with quantum mechanics.  

For us, the most interesting features of birefringence in the Schwarzschild metric are the out-of-plane contributions to trajectory and momentum. First, we have the linearly growing offset -- given by an angle $2\beta $, equation (\ref{angle_beta}) -- between the trajectories of opposite polarisations. Then, the Souriau-Saturnini equations in the Schwarzschild metric become singular far away from the star, a singularity absent in the Kottler metric. Therefore we expect the offset induced by the angle $\gamma$ (\ref{angle_gamma}) to play a  more important role in observations. (The computations with non-vanishing cosmological constant are  complicated and will be presented in a later work.)
In any case, both angles, $\beta $ and $\gamma $ are wavelength dependent and the  offset  must feature a rainbow effect.

Despite the mentioned classical instability, we wonder whether this type of gravitational birefringence is accessible to experimental verification. The upper bound of the 1976 Very Long Baseline Interferometry experiment \cite{vlbi} achieves an upper limit for the birefringence angle $\beta$ (or $\gamma$?) of the order of $10^{-3}\, \arcsec$ for $\lambda \sim 10$ cm. With our formulas, we get for this wavelength $\beta \sim 10^{-6}\, \arcsec$ and $\gamma \sim 10^{-11}\, \arcsec$. After 40 years, the needed accuracy for testing our formulas should not be out of reach.

\vspace{5mm}\noindent
We started this work as a rope team of three. But in September 2018  we lost our friend and guide, Christian Duval. Any error or short coming remaining in this work is ours.

${}$\vspace{6mm}\\
\noindent
{ Acknowledgements:} This work has been carried out thanks to the support of the OCEVU Labex
(ANR-11-LABX-0060) and the A*MIDEX project (ANR-11-IDEX-0001-02) funded
by the "Investissements d'Avenir" French government program managed by
the ANR.

\end{document}